\documentstyle[fancyheadings,prl,
amsmath,amsfonts,isolatin1,aps,prc,epsf,amssymb,epsfig,multicol]{revtex}

\begin{document} 
 
\sloppy 
 
\draft 
 
\bibliographystyle{srt}
 
\title{ Future $e^+ e^-$ Colliders  
Sensitivity to $H b \bar{b}$ Coupling and CP Violation} 
 
\author{ 
V. Braguta$^{a}$ , A. Chalov$^{a}$ ,  
A. Likhoded$^{a,b,}$\footnote{andre@ift.unesp.br}  and    
R. Rosenfeld$^{b,}$\footnote{rosenfel@ift.unesp.br}}

 \address{\it  $^a$ Institute of High Energy Physics} 
\address{\it Protvino, Moscow Region, Russia} 
\vspace{1cm}

\address{\it  $^b$ Instituto de F\'{\i}sica Te\'orica - UNESP} 
\address{\it Rua Pamplona, 145} 
\address{\it 01405-900 S\~{a}o Paulo, SP, Brazil}

\maketitle 
 
\vspace{0.1cm} 
 
\begin{abstract} 
We perform a complete simulation of the process $ e^+ e^- \to b \bar{b} \nu 
\bar{\nu}$, where $\nu$ can be an electron, muon or tau neutrino,  
in the context of a general Higgs coupling to $b$ quarks. 
We parametrize the $H b \bar{b}$ coupling as   
$\frac{m_b}{v} (a + i \gamma_5 b) $. Taking into account interference effects 
between pure Higgs and Standard Model contributions, we find 
that sensitivities of the order of $2$\% and $20$\% can be obtained at  
a future  $e^+ e^-$ collider for 
deviations of the $a$ and $b$ parameters respectively from their Standard  
Model values. Combining our analysis with an 
independent measurement of   
$ \Gamma_{H \to b \bar{b}}$  can provide evidence about the CP nature of 
the Higgs sector. 
\end{abstract} 
 
\vspace{0.2cm} 
 
\noindent 
PACS Categories:    98.80.Cq 
 
\vspace{0.2cm}

%%%%%%%%%%%%%%%%%%%%%%%%%%%%%%%%%%%%%%%%%%%%%%%%%%%%%%%%% 
 
\begin{multicols}{2} 
\relax 
 
\section{Introduction}

The origin of fermion masses and mixings is one of most important issues in 
particle physics. In the Standard Model (SM), the Higgs field alone is  
responsible 
for the electroweak symmetry breaking and mass generation. The SM, however, is 
incomplete and a thorough study of the coupling of the 
remnant Higgs boson (or in fact the lightest (pseudo)scalar boson)  
to fermions can provide hints on the actual mechanism of mass generation.  
 
In a recent paper, we have investigated the possibility of detecting deviations 
from the SM in the Higgs couplings to $\tau$-leptons \cite{us} at future $e^+ 
e^-$ colliders. In this letter, we expand our analysis to the case of the 
Higgs couplings to $b$-quarks, which has a better potential in principle due to 
the larger Yukawa coupling.  
 
For definiteness, we concentrate on the determination of the 
(pseudo)scalar-$b$-$\bar{b}$  
coupling at a Linear Collider with a center-of-mass energy of $\sqrt{s} = 500$ 
GeV and accumulated luminosity of 1 ab$^{-1}$, based on the TESLA design 
\cite{Tesla}. We will assume 
that this particle has already been discovered at the Large Hadron Collider but 
a detailed study of its couplings is missing.  
 
We take into account all relevant contributions to the process  
$ e^+ e^- \to b \bar{b} \nu 
\bar{\nu}$, where $\nu$ can be an electron, muon or tau neutrino. In 
particular, weak gauge boson fusion is the dominant contribution to the subset 
of diagrams containing the Higgs boson 
for $M_H < 180$ GeV at $\sqrt{s} \ge 500$ GeV.

In extensions of the SM with extra scalars and pseudoscalars, the lightest 
spin-0 particle can be an admixture of states without a definite parity.  
Hence, we parametrize the general $H b \bar{b}$ coupling as: 
\begin{equation} 
\frac{m_b}{v} (a + i \gamma_5 b)\;,  
\end{equation} 
where $v = 246$ GeV, $m_b$ is the $b-$quark mass and $a = 1$, $b = 0$  in the  
SM.

\vspace*{-2cm} 
%%%%%%%%%%%%%%%%%%%%%% 
\begin{center} 
\begin{figure} 
\centerline{\epsfxsize=1.3\hsize \epsffile{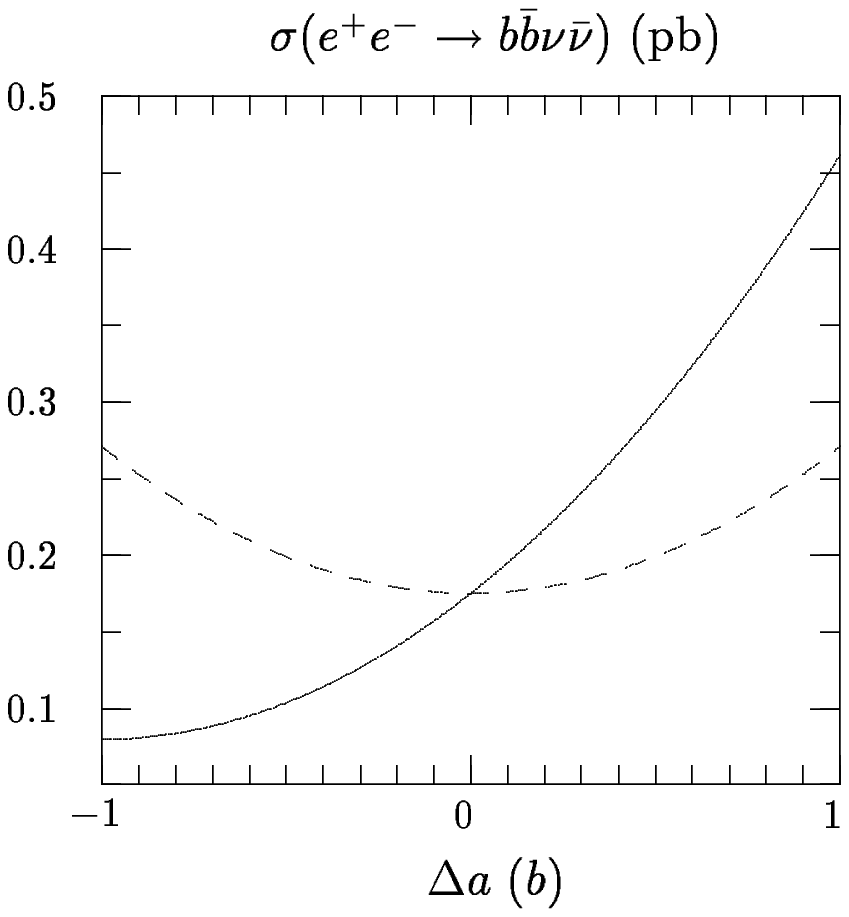}} 
\vskip-8cm 
\caption{The total cross section $e^+e^-\to b \bar{b} \nu \bar\nu$ 
dependence on $\Delta a$ (solid line) and $b$ (dashed line).} 
\label{cross} 
\end{figure} 
\end{center} 
%%%%%%%%%%%%%%%%%%%%%%%%%%%%%%%%%%%%%%%%%%%%%%%%%%%%%%%%%%%%%%%%%%%%%%% 

We will present results  
considering $a$ and $b$ as independent parameters and also for the cases of  
fixed $a=1$, free $b$ and fixed $b=0$, free $a$. We will see that there is a 
region of insensivity around circles in the $a-b$ plane 
since we can't at this level of analysis disentangle the effects of $a$ and $b$.
 
The cross section for the process $ e^+ e^- \to b \bar{b} \nu 
\bar{\nu}$ is sensitive to terms  
proportional to 
$a$, which comes from the interference with non-Higgs contributions, and 
$a^2 $ and $b^2$ from pure Higgs contributions. Therefore,  
we can search for deviations from the SM 
 prediction which could arise, for instance, in supersymmetric models.

The total SM cross section for the process $ e^+ e^- \to b \bar{b} \nu 
\bar{\nu}$ is of the order of $180$ fb for $M_H = 120$ GeV and  
at $\sqrt{s} = 500$ GeV, being dominated by the  
$\nu_e$ final state, because of the many additional diagrams allowed in this 
case. In particular, only in this channel the weak gauge boson fusion  
diagram is allowed and it provides an important contribution. For comparison,  
the process $ e^+ e^- \to H \nu_e 
\bar{\nu}_e$ is of the order of $100$ fb at $\sqrt{s} = 500$ GeV.

In Fig. \ref{cross} we show the dependence of the total cross section,  
summed over the three neutrino species, on 
the parameters $\Delta a \equiv a - 1$ and $b$. The dependence in $b$ is  
symmetric 
since only terms proportional to $b^2$ contribute to the cross section. On the 
other hand, the dependence on $\Delta a$ is asymmetric due to interference with 
Standard Model processes, which leads to the presence of linear terms in  
$\Delta a$.  We should point out that the $\Delta a$ dependence is more 
prominent in the $b \bar{b}$ case than in the previously studied $\tau^+ 
\tau^-$ \cite{us}. Our goal is to see how sensitive the experiments 
performed at the next generation of $e^+e^-$ colliders will be in the  
determination of these parameters.  
 
%\vskip-2cm 
 
\section{Analysis and Results} 
 
We performed our Monte Carlo simulation by generating observables 
represented as series in the $a$ and $b$ couplings multiplied by kinematical 
factors:  
\begin{eqnarray} 
\frac{d\sigma }{d {\cal O} } &=& A_0 + a\cdot A_1  
+ a^2\cdot A_2+ ab\cdot A_3+ \\ \nonumber  
&& b\cdot A_4+ b^2\cdot A_5 \mbox{\dots } 
\end{eqnarray} 
where ${\cal O}$ is any observable and the $A_i$ terms are purely kinematical  
structures which do not contain 
any $a$ and $b$ dependence and results from the amplitude squaring and phase 
space integration. 
These $A_i$ structures are the subject of the Monte Carlo simulation and 
the $a$ and $b$ couplings can be varied without the 
necessity to re-simulating the data for each $(a,b)$ point. In our particular 
case, $A_3 =  A_4 = 0$.

The event sample reproducing the expected statistics at TESLA was generated 
using our Monte Carlo package while the detector response was simulated with 
the code {\tt SIMDET} version 3.01\cite{simdet}. We assume an efficiency  
for $b-$jet pair reconstruction of $ \varepsilon_{bb}=56$ \%,  
which is based on the $b$-tag 
algorithm, as assumed in ref. \cite{lcnote}. In our simulations we used $M_H = 
120$ GeV.

\begin{center} 
\begin{figure} 
\centerline{\epsfxsize=1\hsize \epsffile{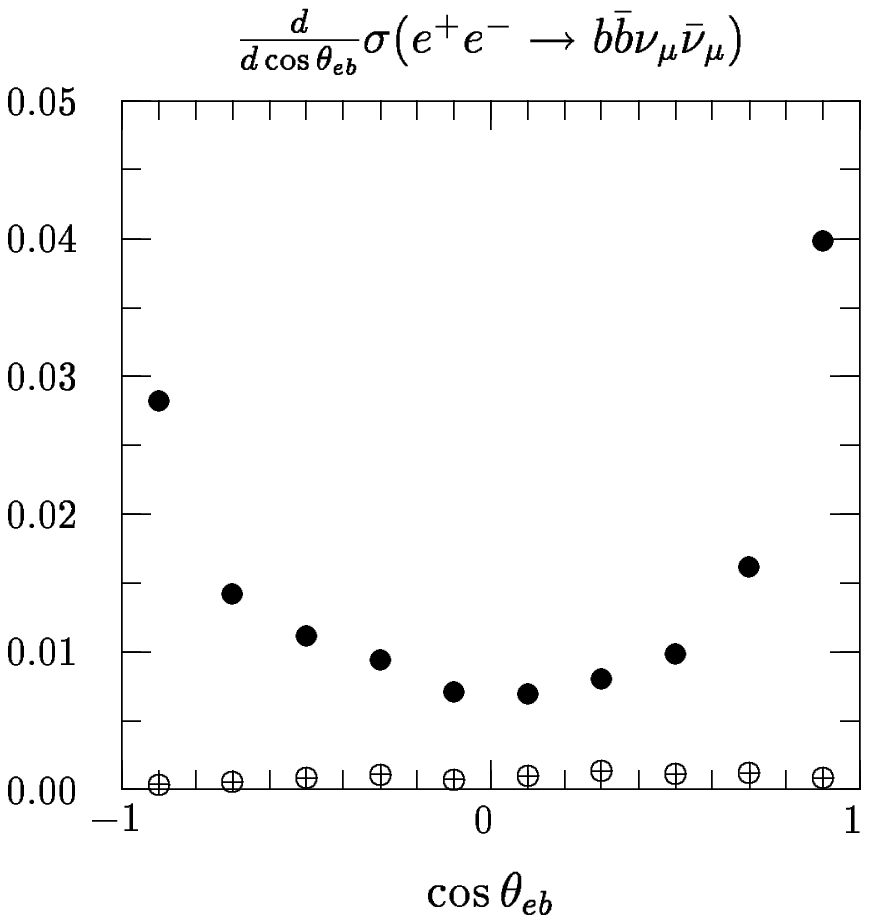} 
\hskip-4cm  
\epsfxsize=1\hsize \epsffile{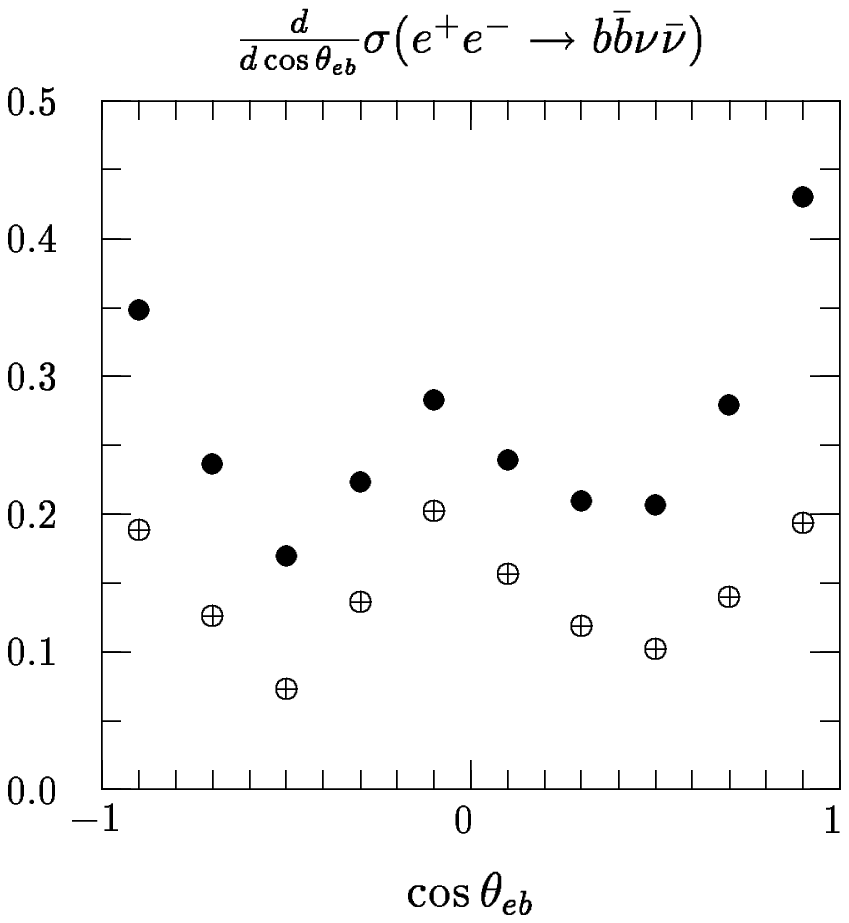}} 
\vskip-6.5cm 
\caption{  
The differential cross section over $\cos\theta_{eb}$ (in pb)  
for the process $e^+e^-\to b \bar{b} \nu_\mu \bar{\nu_\mu}$ and 
 $e^+e^-\to b \bar{b} \nu \bar{\nu}$, summed over the three neutrinos.  
Solid circles are the full  
SM result while crossed circles are the contribution from Higgs only  
(including interference effects). } 
\label{distribution} 
\end{figure} 
\end{center}

In Figure \ref{distribution} we show, for comparison purposes, the  
differential distribution in $\cos\theta_{eb}$, where $\theta_{eb}$
is the scattering angle between the $b$-jet and initial beam directions, 
for the total SM contribution and for the Higgs contribution only  
(including interference with SM). To illustrate the importance of the  
$\nu_e$ final state, we plot both the contribution of only the  
$\nu_\mu$ final state (which is basically the same as $\nu_\tau$ final state)  
and the total contribution from the three neutrinos. We can see that in the 
first case the Higgs contribution is small but in the total contribution it is 
comparable to the full SM result.

%\vspace*{-2cm} 
%%%%%%%%%%%%%%%%%%%%%%%%%%%%%%%%%%%%%%%%%%%%% 
%%%%%%%%%%%%%%%%%%%%%%%%%%%%%%%%%%%%%%%%%%%%%%%%%% 

\vskip-2.5cm 
%%%%%%%%%%%%%%%%%%%%%%%%%%%%%%%%%%%%%%%%%%%%%%%%%%%%%%%%%%%%%%%%%% 
\begin{figure} 
\centerline{\epsfxsize=1\hsize \epsffile{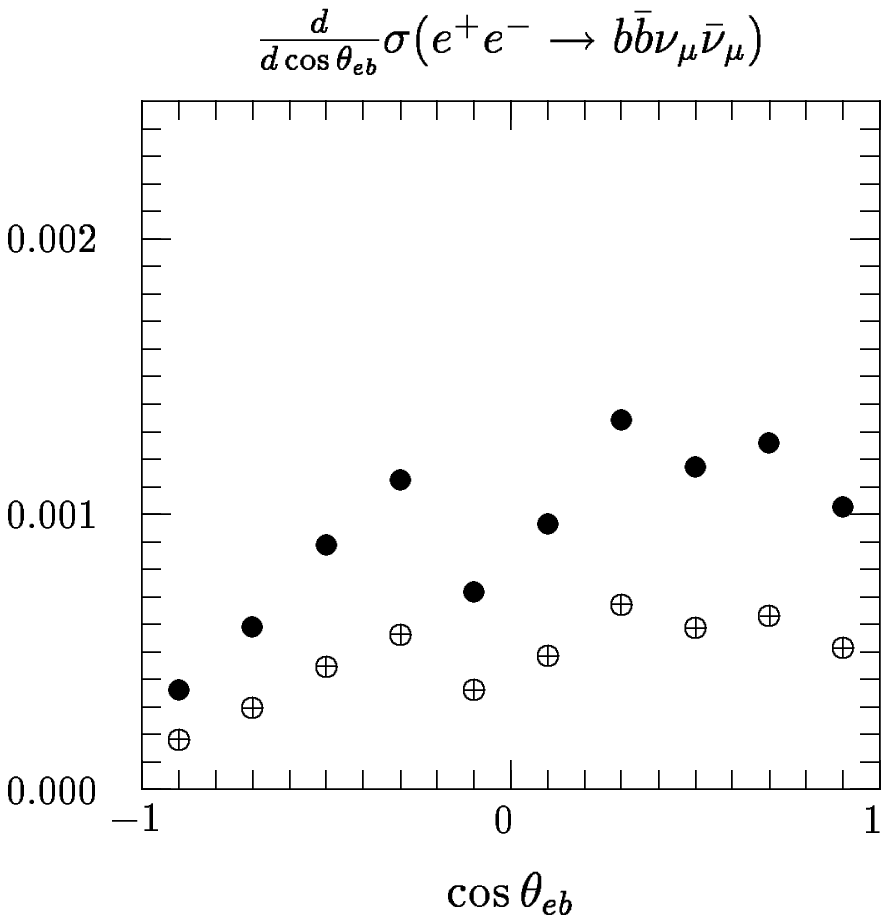} 
\hskip-4cm  
\epsfxsize=1\hsize \epsffile{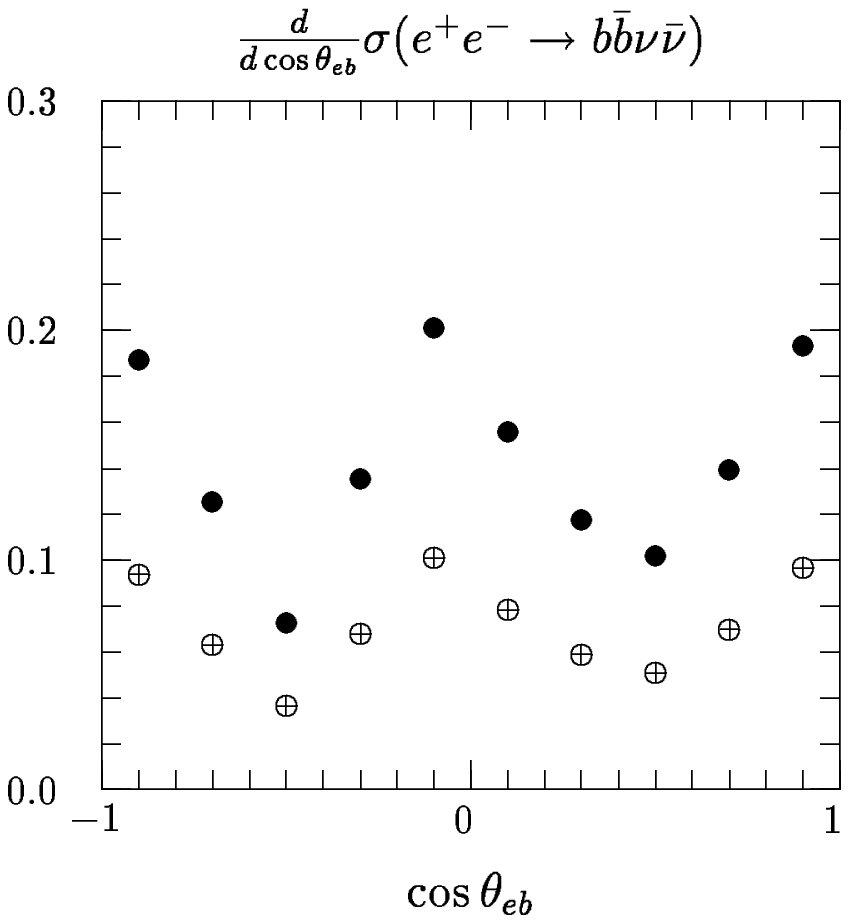}} 
\vskip-6cm 
\caption{ 
Contribution of Higgs diagrams to the differential $\cos\theta_{eb}$  
distribution in the  $e^+e^-\to b \bar{b} \nu \bar\nu$ process for 
Standard Model ($a=1, b=0$, black dots) and $a=0.5, b=0.5$ (crossed dots).} 
\label{ab} 
\end{figure} 
%%%%%%%%%%%%%%%%%%%%%%%%%%%%%%%%%%%%%%%%%%%%%% 

%\vspace*{-1cm} 
In order to demonstrate the effect of different values of the parameters $a$ and 
$b$, we show in Figure \ref{ab} the $\cos\theta_{eb}$ distribution arising from
the Higgs contribution for the SM  
($a=1$, $b=0$) 
compared to the case with $a = b = 0.5$, with only $\nu_\mu$ neutrinos and with 
the three neutrinos.  We see that the shapes are very similar, as expected, but  
the levels can be noticeable different. 
 
As for the possible contribution from background processes, like 
$e^+e^-\to e^+e^- ZZ\to e^+e^- b \bar{b} \nu\bar\nu$ (with $e$-pair lost), 
$e^+e^-\to \nu\bar\nu W^+W^- \to \nu\bar\nu b \bar{b}\nu\bar\nu$,  
$e^+e^-\to ZZZ\to b \bar{b}\nu\bar\nu\nu\bar\nu$, etc., the cross sections 
of these processes are either small, or they can be significantly 
supressed down to levels of $0.2$ fb \cite{lcnote}.

Another important aspect is the assumption about the detector performance  
and possible sources of the systematic uncertaintes. We  
include the anticipated  systematic errors of 0.5\% in the luminosity  
measurement, 
1\%  in the acceptance determination, 1\% in the branching ratios, and 
1\% in the background substraction, and assume the Gaussian nature of the  
sytematics. To place bounds on the $H b \bar{b}$ couplings, we  use a standard  
$\chi^2$-criterion to analyse the events. After various kinematical
distributions were examined, we found that the  
most strict bounds are achieved from $\cos\theta_{eb}$ distribution
by dividing the distribution event samples into 10 bins.  
The experimental error $\Delta 
\sigma^{exp}_i$ for the $i^{th}$ bin is given by: 
\begin{equation} 
\Delta\sigma^{exp}_i = \sigma^{SM}_i \sqrt{\delta^2_{syst} + \delta^2_{stat} } 
\end{equation} 
where  
\begin{equation} 
\delta_{stat}  = \frac{1}{\sqrt{\sigma^{SM}_i \varepsilon_{bb} \int  
{\cal L} dt  
}} 
\end{equation} 
and $\delta_{syst}^2$ is the sum in quadrature of the systematic uncertainties 
mentioned above.

\begin{figure} 
\centerline{\epsfxsize=1.5\hsize \epsffile{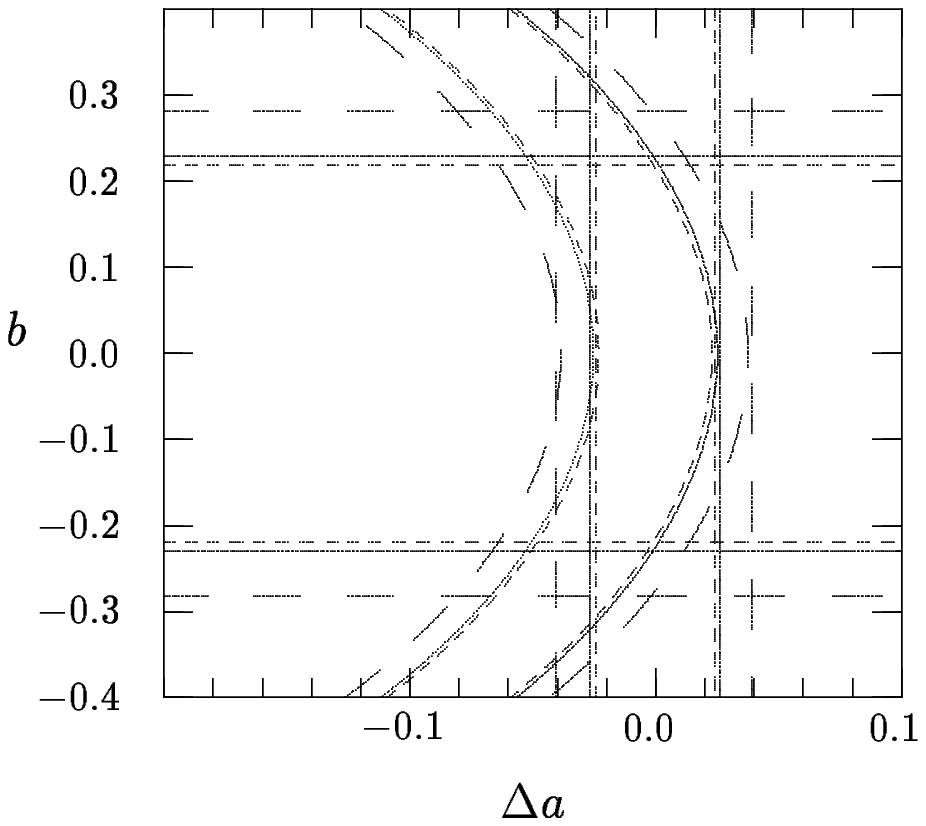}} 
\vskip-10cm 
\caption{ 
The allowed regions for $\Delta a$ and $b$ parameters  
at 95\% confidence level for ${\cal L} = 100$ fb$^{-1}$ (long-dashed lines), 
$1$ ab$^{-1}$ (solid lines) and $10$ ab$^{-1}$ (short-dashed lines). 
The area between the circles is the allowed region for independent  
$\Delta a$ and $b$. 
The horizontal bands are the allowed region for the $b$ 
parameter keeping $\Delta a = 0$. The vertical bands  are the  
the allowed region for the $\Delta a$ 
parameter keeping $b=0$. 
} 
\label{contour} 
\end{figure}

In Fig. \ref{contour}, we present our final results for a TESLA-like 
environment \cite{Tesla} with a center-of-mass energy of 500 GeV 
and for $M_H= 120$ GeV. 
 
We investigated three possible scenarios for the luminosities: 100 fb$^{-1}$,   
1 ab$^{-1}$ and 10 ab$^{-1}$. 
The allowed region for independent $\Delta a$ and $b$ parameters  
at 95\% confidence level is the area between the circles. 
The horizontal bands are the allowed region for the $b$ 
parameter keeping $a=1$. The vertical bands  are  
the allowed region for the $\Delta a$ 
parameter keeping $b=0$.

\vspace*{-2cm} 
%%%%%%%%%%%%%%%%%%%%%%%%%%%%%%%%%%%%%%%%%%%%%%%% 
%%%%%%%%%%%%%%%%%%%%%%%%%%%%%%%%%%%%%%%%% 
%\vspace*{-1cm} 
 
The bounds that can be obtained at 95\% confidence level  are: 
\begin{eqnarray} 
-0.041 &\leq& \Delta a\leq 0.039\; \mbox{ for ${\cal L} = 100$ fb$^{-1}$}; \\ 
\nonumber 
-0.026 &\leq& \Delta a\leq 0.027\; \mbox{ for ${\cal L} = 1$ ab$^{-1}$}; \\ 
-0.024 &\leq& \Delta a\leq 0.024\; \mbox{ for ${\cal L} = 10$ ab$^{-1}$}. \nonumber 
\end{eqnarray} 
for the case of $b=0$ and free $\Delta a$ and 
\begin{eqnarray} 
-0.28 &\leq& b\leq 0.28\; \mbox{ for ${\cal L} = 100$ fb$^{-1}$}; \\\nonumber 
-0.23 &\leq& b\leq 0.23\; \mbox{ for ${\cal L} = 1$ ab$^{-1}$}; \\ 
-0.22 &\leq& b\leq 0.22\; \mbox{ for ${\cal L} = 10$ ab$^{-1}$}.\nonumber 
\end{eqnarray} 
for the case of $\Delta a=0$ and free $b$. 
 
These results are up to an order of magnitude better than the limits obtained 
in a similar manner for $\tau-$leptons, mainly because of the larger Yukawa 
coupling in case of $b-$quarks and higher sensitivity of the process.  
 
These results can be roughly scaled for moderate variations in the Higgs boson  
mass around $120$ GeV by multiplying the bounds by a factor $(M_H/120 \mbox{ 
GeV})^2$.

\section{Conclusions} 
 
We have performed a  complete analysis of the sensitivity to new $H b \bar{b}$  
couplings from the process $ e^+ e^- \to b \bar{b} \nu 
\bar{\nu}$ at the next generation of linear colliders. These new couplings are 
predicted by many extensions of the Standard Model. We showed that forthcoming 
experiments will be able to probe deviations of $H b \bar{b}$ coupling. 
The weak gauge boson fusion process is instrumental for achieving such a 
precision.  
For a TESLA-like environment, we are able to constrain the couplings at the 
level of a few percent for the $a$ parameter (for fixed $b$) and tens of  
percent for the $b$ parameter (for fixed $a$). These results are comparable to 
the study performed in \cite{Tesla}, where a global fit analysis for ${\cal L} 
= 500$ fb$^{-1}$ and $\sqrt{s} = 500$ GeV has resulted in a relative accuracy of 
$2.2$\% in the $g_{Hbb}$ Yukawa coupling.

We would like to comment some aspects of the future measurements. Let us 
assume that the Higgs data anticipated from the new collider experiments will 
reveal a deviation  
from the SM predictions. 
In addition, suppose that one has an independent  measurement 
of the  partial width $\Gamma_{H\to b \bar b}$ (for instance, 
from on-mass-shell Higgs production in  a muonic collider).  
It easy to see that, in our parametrization, 
$\Gamma_{H\to b \bar b}\sim (a^2 +b^2)$, 
while the observables we studied have the following parameter dependence 
\begin{displaymath} 
\frac{d\sigma }{d {\cal O} } = A_0 + a\cdot A_1  
+ a^2\cdot A_2+ b^2\cdot A_3\;.  
\end{displaymath} 
Combining our results and those from $\Gamma_{H\to b \bar b}$ one can separate 
$a$ and $b$ contributions and obtain an explicit indication  
of CP violation in the Higgs sector.

\section*{Acknowledgments} 
 
The work of A. Likhoded was partially funded by a Fapesp grant  
2001/06391-4. The work of A. Chalov and V.Braguta is partially supported 
by the Russian Foundation for Basic Research, grants 99-02-16558 and 00-15-96645, 
Russian Education Ministry, 
grant RF~E00-33-062, and CRDF grant MO-011-0. 
R. Rosenfeld would like to thank Fapesp and CNPq for partial 
financial support. The authors would like to thank A. Belyaev for
valuable remarks.

% Journal and other miscellaneous abbreviations for references 
\def \arnps#1#2#3{Ann.\ Rev.\ Nucl.\ Part.\ Sci.\ {\bf#1} (#3) #2} 
\def \art{and references therein} 
\def \cmts#1#2#3{Comments on Nucl.\ Part.\ Phys.\ {\bf#1} (#3) #2} 
\def \cn{Collaboration} 
\def \cp89{{\it CP Violation,} edited by C. Jarlskog (World Scientific, 
Singapore, 1989)} 
\def \econf#1#2#3{Electronic Conference Proceedings {\bf#1}, #2 (#3)} 
\def \efi{Enrico Fermi Institute Report No.\ } 
\def \epjc#1#2#3{Eur.\ Phys.\ J. C {\bf#1} (#3) #2} 
\def \f79{{\it Proceedings of the 1979 International Symposium on Lepton and 
Photon Interactions at High Energies,} Fermilab, August 23-29, 1979, ed. by 
T. B. W. Kirk and H. D. I. Abarbanel (Fermi National Accelerator Laboratory, 
Batavia, IL, 1979} 
\def \hb87{{\it Proceeding of the 1987 International Symposium on Lepton and 
Photon Interactions at High Energies,} Hamburg, 1987, ed. by W. Bartel 
and R. R\"uckl (Nucl.\ Phys.\ B, Proc.\ Suppl., vol.\ 3) (North-Holland, 
Amsterdam, 1988)} 
\def \ib{{\it ibid.}~} 
\def \ibj#1#2#3{~{\bf#1} (#3) #2} 
\def \ichep72{{\it Proceedings of the XVI International Conference on High 
Energy Physics}, Chicago and Batavia, Illinois, Sept. 6 -- 13, 1972, 
edited by J. D. Jackson, A. Roberts, and R. Donaldson (Fermilab, Batavia, 
IL, 1972)} 
\def \ijmpa#1#2#3{Int.\ J.\ Mod.\ Phys.\ A {\bf#1} (#3) #2} 
\def \ite{{\it et al.}} 
\def \jhep#1#2#3{JHEP {\bf#1} (#3) #2} 
\def \jpb#1#2#3{J.\ Phys.\ B {\bf#1} (#3) #2} 
\def \jpg#1#2#3{J.\ Phys.\ G {\bf#1} (#3) #2} 
\def \mpla#1#2#3{Mod.\ Phys.\ Lett.\ A {\bf#1} (#3) #2} 
\def \nat#1#2#3{Nature {\bf#1} (#3) #2} 
\def \nc#1#2#3{Nuovo Cim.\ {\bf#1} (#3) #2} 
\def \nima#1#2#3{Nucl.\ Instr.\ Meth. A {\bf#1} (#3) #2} 
\def \npb#1#2#3{Nucl.\ Phys.\ B {\bf#1} (#3) #2} 
\def \npps#1#2#3{Nucl.\ Phys.\ Proc.\ Suppl.\ {\bf#1} (#3) #2} 
\def \npbps#1#2#3{Nucl.\ Phys.\ B Proc.\ Suppl.\ {\bf#1} (#3) #2} 
\def \PDG{Particle Data Group, D. E. Groom \ite, \epjc{15}{1}{2000}} 
\def \pisma#1#2#3#4{Pis'ma Zh.\ Eksp.\ Teor.\ Fiz.\ {\bf#1} (#3) #2 [JETP 
Lett.\ {\bf#1} (#3) #4]} 
\def \pl#1#2#3{Phys.\ Lett.\ {\bf#1} (#3) #2} 
\def \pla#1#2#3{Phys.\ Lett.\ A {\bf#1} (#3) #2} 
\def \plb#1#2#3{Phys.\ Lett.\ B {\bf#1} (#3) #2} 
\def \pr#1#2#3{Phys.\ Rev.\ {\bf#1} (#3) #2} 
\def \prc#1#2#3{Phys.\ Rev.\ C {\bf#1} (#3) #2} 
\def \prd#1#2#3{Phys.\ Rev.\ D {\bf#1} (#3) #2} 
\def \prl#1#2#3{Phys.\ Rev.\ Lett.\ {\bf#1} (#3) #2} 
\def \prp#1#2#3{Phys.\ Rep.\ {\bf#1} (#3) #2} 
\def \ptp#1#2#3{Prog.\ Theor.\ Phys.\ {\bf#1} (#3) #2} 
\def \ppn#1#2#3{Prog.\ Part.\ Nucl.\ Phys.\ {\bf#1} (#3) #2} 
\def \rmp#1#2#3{Rev.\ Mod.\ Phys.\ {\bf#1} (#3) #2} 
\def \rp#1{~~~~~\ldots\ldots{\rm rp~}{#1}~~~~~} 
\def \si90{25th International Conference on High Energy Physics, Singapore, 
Aug. 2-8, 1990} 
\def \zpc#1#2#3{Zeit.\ Phys.\ C {\bf#1} (#3) #2} 
\def \zpd#1#2#3{Zeit.\ Phys.\ D {\bf#1} (#3) #2}

\end{multicols} 
 
\end{document}